\documentclass[aps,prc,reprint,superscriptaddress,nofootinbib]{revtex4-2}
\usepackage{amsmath}
\usepackage{amssymb}
\usepackage{graphicx}% Include figure files
\usepackage{dcolumn}% Align table columns on decimal point
\usepackage{bm}% bold math
\usepackage{braket}
\usepackage{mathrsfs}
\usepackage{slashed}

\begin{document}

%\preprint{}

\title{
Toward experimental determination of spin entanglement of nucleon pairs
}

\author{Dong Bai}
\email{dbai@hhu.edu.cn}
\affiliation{College of Mechanics and Engineering Science, Hohai University, Nanjing 211100, China
}%

%\author{Zhongzhou Ren}
%\email{zren@tongji.edu.cn}
%\affiliation{School of Physics Science and Engineering, Tongji University, Shanghai 200092, China}%
%\affiliation{Key Laboratory of Advanced Micro-Structure Materials, Ministry of Education, Shanghai 200092, China}

%\date{\today}

\begin{abstract}

%The problem of experimental determination of the amount of entanglement in an abstract two-qubit pure state
%has been studied by [J.\ M.\ G.\ Sancho and S.\ F.\ Huelga, Phys.\ Rev.\ A {\bf 61}, 042303 (2000)].
 
Nuclear entanglement is a flagship in the interdisciplinary direction of nuclear physics and quantum information science.
Spin entanglement, a special kind of nuclear entanglement, is ubiquitous in nuclear structures and dynamics.
Based on the idea of quantum state tomography,
the problem of experimental determination of spin entanglement of two-nucleon pure states is studied directly
within the scope of nuclear physics. 
It is shown that the amount of spin entanglement can be obtained by measuring three spin polarizations of one nucleon in polarization experiments.
The errors from imperfect preparation of nucleon pairs are also analyzed.
This work not only complements the existing literature of nuclear entanglement
but also provides new opportunities for nuclear physics with polarized particles. 

%%It is being investigated intensively from the theoretical side.
%How to measure it in nuclear experiments?
%We study this problem for spin entanglement of nucleon pairs with unknown spin wave functions.
%It is shown that
%the degree of spin entanglement can be determined experimentally 
%by measuring a small number of spin polarizations,
%which seems feasible with current or future technologies.
 
\end{abstract}

\maketitle

%\tableofcontents

\section{Introduction}

Nuclear entanglement is the latest
frontier for the century-long exploration of the quantum nature of nuclear systems,
%In literature, nuclear quantumness has been constantly investigated from the aspects of 
%the quantal behaviors of nuclear observables (e.g., shell-model explanation of nuclear magic numbers) and the dual relations between hadrons and nuclear fields (e.g., Yukawa's meson-exchange picture of nuclear forces).
%%Not directly related to the above two aspects,
%On the other hand, nuclear entanglement 
revealing a previously unexplored aspect of nuclear quantumness \cite{Kwasniewicz:2013cqa,Kanada-Enyo:2015ncq,Kanada-Enyo:2015kyo,Legeza:2015fja,Gorton:2018,Johnson:2019,Kruppa:2020rfa,Kovacs:2021yme,Kwasniewicz:2016,Kwasniewicz:2017dbc,Gorton:2018,Johnson:2019,Beane:2018oxh,Robin:2020aeh,Low:2021ufv,Beane:2021zvo,Jafarizadeh:2022kcq,Kovacs:2021yme,Pazy:2022mmg,Tichai:2022bxr,Bai:2022hfv,Liu:2022grf,Johnson:2022mzk,Bai:2023rkc,Gu:2023aoc,Bulgac:2022cjg,Bulgac:2022ygo,Sun:2023nzj,Bai:2023tey,Miller:2023ujx,Miller:2023snw,Hengstenberg:2023ryt,Perez-Obiol:2023wdz,Kirchner:2023,Kou:2023knx}.
It refers to a novel kind of nonclassical correlation that is ubiquitous in nuclear systems
but would be absent if nuclear physics obeyed classical laws.
%understanding nuclear entanglement needs the interdisciplinary efforts from nuclear physics and quantum information science.
The study of nuclear entanglement is inspired by recent progress in quantum information science and can be regarded a systematic extension of traditional approaches to nuclear correlations.  
It also provides valuable guidance to solve nuclear many-body problems on classical and quantum computers. 

%Nuclear entanglement were first studied in the 1970s.
In 1976, Lamehi-Rachti and Mittig tested Bell's inequalities by measuring spin correlations in low-energy proton-proton scattering \cite{Lamehi-Rachti:1976}.
In 2006, Sakai \emph{et al.}\ gave a new nuclear Bell test by using proton pairs coming from the $p(d,{}^2\text{He})n$ reaction,
improving Lamehi-Rachti and Mittig's analysis in several key aspects \cite{Sakai:2006} (see also Ref.~\cite{Polachic:2004}).
Given the close connection between Bell's inequality violation and quantum entanglement,
these studies actually pioneer the direction of nuclear entanglement. 
%These experiments establish the Bell's inequality violation in nuclear systems.
%Given the close relation between entanglement and Bell's inequality violation,
%it is fair to say that they actually
% pioneer the studies of nuclear entanglement.
%The meaning of these studies cannot be overstated.
Although widely recognized and highly cited by the quantum information community, these studies do not attract much attention from the nuclear physics community,
since both nuclear Bell tests and nuclear entanglement have remained outside the main stream of nuclear physics for decades.
%The above works failed to attract attention from nuclear physics community for decades.

The situation is changing now.
In the last few years, nuclear entanglement has been studied by several groups from the theoretical side
and a number of interesting theoretical results are obtained for nuclear structure and reaction problems \cite{Kwasniewicz:2013cqa,Kanada-Enyo:2015ncq,Kanada-Enyo:2015kyo,Legeza:2015fja,Gorton:2018,Johnson:2019,Kruppa:2020rfa,Kovacs:2021yme,Kwasniewicz:2016,Kwasniewicz:2017dbc,Gorton:2018,Johnson:2019,Beane:2018oxh,Robin:2020aeh,Low:2021ufv,Beane:2021zvo,Jafarizadeh:2022kcq,Kovacs:2021yme,Pazy:2022mmg,Tichai:2022bxr,Bai:2022hfv,Liu:2022grf,Johnson:2022mzk,Bai:2023rkc,Gu:2023aoc,Bulgac:2022cjg,Bulgac:2022ygo,Sun:2023nzj,Bai:2023tey,Miller:2023ujx,Miller:2023snw,Hengstenberg:2023ryt,Perez-Obiol:2023wdz,Kirchner:2023,Kou:2023knx}. 
%Several calculations have been made to study nuclear entanglement in nuclear structure and reaction problems.
%In Refs.~\cite{},
%spin entanglement of nucleon-nucleon and nucleon-nucleus scattering is studied, as well as its connections to the Wigner and Schr\"odinger symmetries.
%In Refs.~\cite{},
%the scaling behavior of entanglement entropy is investigated for nuclear systems,
%and the volume law is more favored than the conventional area law.
%In Refs.~\cite{},
%the entanglement rearrangement is studied by modifying in nuclear many-body calculations.
Given these theoretical achievements, it may be timely to also investigate how to determine the amount of nuclear entanglement experimentally. 
%Besides the old connections to experimental Bell tests in nuclear physics, another interesting problem is how to measure nuclear entanglement experimentally. 
As far as I know, this problem has not been discussed publicly in low-energy nuclear physics.
In quantum information science, the experimental determination of the amount of quantum entanglement has been studied from the formal aspects by Refs.~\cite{Sancho:2000,Walborn:2006,Walborn:2007,Fei:2009,Li:2013} for two-qubit pure states.
A straightforward protocol is based on quantum state tomography (QST) \cite{Paris:2004}, reconstructing the reduced density matrix from a couple of local measurements on one of the qubits
and then evaluating the entanglement measure correspondingly (see Sec.~\ref{SC} for a detailed description).
It is shown by Ref.~\cite{Sancho:2000} that the QST protocol is actually optimal under the following three conditions: 
(1) only one pair of qubits can be measured at each time, (2) there are no auxiliary qubits available and only projective measurements are allowed in experiments,
and 
(3) the measurement protocol remains unchanged in experiments.
Later on, it was found that, if the second condition is relaxed, 
an improved protocol can be designed with a single local measurement,
provided the availability of an auxiliary two-qubit system in the same state as the original one \cite{Walborn:2006,Walborn:2007}. 
Both protocols have been realized successfully in photonic experiments \cite{Walborn:2006,Walborn:2007,White:1999}.
However, the experimental techniques for nucleon manipulation are generally different from those for photon manipulation.
Some projective operators realizable in photonic experiments may be too complicated to be implemented in nuclear experiments.
A detailed discussion will be given in Sec.~\ref{Concur},
related to the second protocol mentioned above.
Therefore, it is necessary to work out the experimental protocol directly within the scope of low-energy nuclear physics.
%While similar problems have been studied in photonic, atomic, condensed matter physics,
%not all those studies can be applied directly to low-energy nuclear physics,

%For example, an experimental setup is reported in Ref.~\cite{} to detect 
%polarization entanglement in photon pairs experimentally,
%which is made up of 
%nonlinear crystals, wave plates, polarizing beam splitters, and other optical devices. 
%It would be difficult to 
%adapt this setup to nuclear experiments without essential modifications.
%Therefore, it is necessary to study these problems directly in the context of nuclear physics.

As a first attempt, I study the experimental determination of spin entanglement of two-nucleon pure states with unknown spin wave functions.
%The primary goal is to convince the community that 
Explicitly, the neutron-proton pair is under consideration.
In principle, the discussions could also be extended to proton-proton and neutron-neutron pairs, 
after taking into consideration the additional complexity due to identical particles.
Spin entanglement is a special example of nuclear entanglement involving only spin degrees of freedom.
It is ubiquitous in nuclear systems as nucleons are spin-1/2 particles and nuclear forces are spin dependent.
These two nucleons could be produced by nucleon-nucleon scattering, dinucleon decays, knockout reactions, etc,
and all these processes are named as entanglement sources for convenience.
It is assumed that an ideal entanglement source can produce an infinite number of identical two-nucleon pure states,
upon which the experimental measurements are carried out.
%Concerning the problem of the experimental measurability of spin entanglement,
%it is not hard to convince oneself that the answer is yes.
%In principle, the two-nucleon spin wave function lives in a four-dimensional Hilbert space.
%Explicitly, I will show that how to determine the spin concurrence, the widely used entanglement measure, from a small number of spin correlation functions. 
%As a general quantum phenomenon, entanglement has been the central concept of quantum information science for more than three decades.
The reminder of this work is organized as follows.
In Sec.~\ref{SC},
the basic properties of reduced density matrix and spin entanglement are discussed
with the emphasis on an entanglement measure called concurrence and its representations in terms of spin polarizations.
In Sec.~\ref{SCM},
the possible experimental setups to determine spin polarizations are discussed,
from which the experimental estimation of concurrence could be obtained.
In Sec.~\ref{Concl},
conclusions are drawn.

\section{Spin Entanglement}
\label{SC}

\subsection{Reduced density matrix}

 \begin{figure}

\centering
  \includegraphics[width=0.35\linewidth]{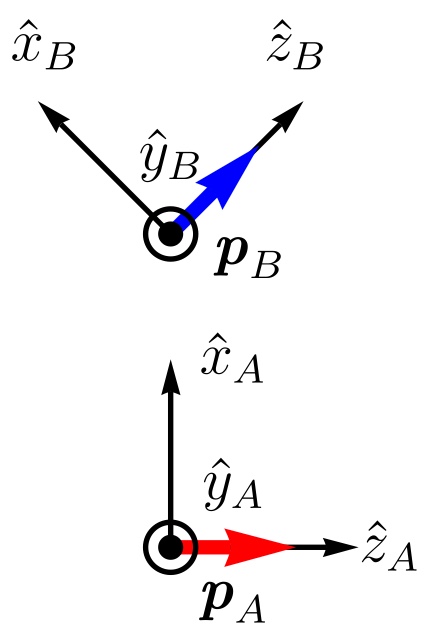}
  \caption{
The coordinate system for the nucleon pair in the laboratory frame,
with $\bm{p}_i$ $(i=A,B)$
being the nucleon momentum and $\hat{x}_i$, $\hat{y}_i$, $\hat{z}_i$
being three unit vectors given by Eq.~\eqref{UV}.
In this coordinate system, the two nucleons move along the $\hat{z}_A$ and $\hat{z}_B$ axes in the $\hat{x}_A$-$\hat{z}_A$ plane (equivalently, the $\hat{x}_B$-$\hat{z}_B$ plane),
while $\hat{y}_A$ and $\hat{y}_B$ point to the same direction perpendicular to the scattering plane.
}

  \label{Madison}
       
\end{figure}

Suppose there is an entangled neutron-proton pair emitted from an entanglement source.
%The two nucleons then move apart and become distant from each other after some time, with the separation distance being much larger than the typical interaction range of strong interactions.
%Therefore, in our discussions, two protons or two neutrons can also be safely regarded as distinguishable particles due to the cluster decomposition principle, in addition to the case of one proton and one neutron.
The momenta of the two nucleons (labeled as $A$ and $B$) are given by $\bm{p}_A$ and $\bm{p}_B$ in the laboratory frame.
For the later convenience, the following coordinate system is introduced for each nucleon $(i=A,B)$
with the orthonormal vectors given by
\begin{align}
\hat{z}_i=\frac{\bm{p}_i}{|\bm{p}_i|},\quad\ \ \hat{y}_i=\frac{\bm{p}_A\times\bm{p}_B}{|\bm{p}_A\times\bm{p}_B|},\quad\ \ \hat{x}_i=\hat{y}_i\times\hat{z}_i.
\label{UV}
\end{align}
See also Fig.~\ref{Madison}.
In this coordinate system, the two nucleons move along $\hat{z}_A$ and $\hat{z}_B$, respectively,
and $\hat{y}_A$ and $\hat{y}_B$ point to the same direction perpendicular to the scattering plane spanned by $\bm{p}_A$ and $\bm{p}_B$.
For either set of $\{\hat{x}_i,\hat{y}_i,\hat{z}_i\}$,
 the corresponding Pauli matrices are denoted by
$\{\sigma_{\hat{x}_i},\sigma_{\hat{y}_i},\sigma_{\hat{z}_i}\}$.

Without loss of generality, the following wave function is considered for the two nucleons,
\begin{align}
&\ket{\Psi_{AB}}=\ket{\bm{p}_A\bm{p}_B}\otimes\ket{\Psi_{AB}^{\text{spin}}},\label{PsiAB}\\
&\ket{\Psi_{AB}^{\text{spin}}}=\alpha_{00}\ket{0_A0_B}+\alpha_{01}\ket{0_A1_B}\nonumber\\
&\ \ \ \ \ \ \ \ \ \,+\alpha_{10}\ket{1_A0_B}+\alpha_{11}\ket{1_A1_B}\label{PsiABspin}
\end{align}
with $\ket{0_i}$ and $\ket{1_i}$ being the eigenstates of $\sigma_{\hat{z}_i}$ satisfying $\sigma_{\hat{z}_i}\ket{0_i}=\ket{0_i}$ and $\sigma_{\hat{z}_i}\ket{1_i}=-\ket{1_i}$,
and $\alpha_{00}$, $\alpha_{01}$, $\alpha_{10}$, $\alpha_{11}$ are the expansion coefficients satisfying 
$|\alpha_{00}|^2+|\alpha_{01}|^2+|\alpha_{10}|^2+|\alpha_{11}|^2=1$.
The two-nucleon wave function $\ket{\Psi_{AB}}$ in the form of Eq.~\eqref{PsiAB} can be produced naturally by the nucleon-nucleon scattering \cite{Bai:2023tey}.
The reduced density matrix $\rho^\text{spin}_A=\text{Tr}_B\ket{\Psi^\text{spin}_{AB}}\bra{\Psi^\text{spin}_{AB}}$ ($\rho^\text{spin}_B=\text{Tr}_A\ket{\Psi^\text{spin}_{AB}}\bra{\Psi^\text{spin}_{AB}}$) is obtained from the spin wave function $\ket{\Psi_{AB}^{\text{spin}}}$ by tracing out $B$ ($A$).
Explicitly, $\rho_{A}^\text{spin}$ is given by
\begin{align}
\rho^\text{spin}_A=\left[
\begin{array}{cc}
|\alpha_{00}|^2+|\alpha_{01}|^2 & \alpha_{00}^*\alpha_{10}+\alpha_{01}^*\alpha_{11} \\
\alpha_{00}\alpha_{10}^*+\alpha_{01}\alpha_{11}^* & |\alpha_{10}|^2+|\alpha_{11}|^2
\end{array}\right],
\end{align}
and $\rho_{B}^\text{spin}$ can be derived in a similar way.
As a $2\times2$ Hermitian matrix, $\rho^\text{spin}_i$ $(i=A,B)$ can also be parametrized by
\begin{align}
\rho^\text{spin}_i
=&\frac{1}{2}\!\left(1
+\braket{\sigma_{\hat{x}_i}}_{\!i}\sigma_{\hat{x}_i}
+\braket{\sigma_{\hat{y}_i}}_{\!i}\sigma_{\hat{y}_i}
+\braket{\sigma_{\hat{z}_i}}_{\!i}\sigma_{\hat{z}_i}\right)\label{rhospina}\\
=&
\frac{1}{2}\left[
\begin{array}{cc}
1+\braket{\sigma_{\hat{z}_i}}_{\!i} & \braket{\sigma_{\hat{x}_i}}_{\!i}-i\braket{\sigma_{\hat{y}_i}}_{\!i}\\
\braket{\sigma_{\hat{x}_i}}_{\!i}+i\braket{\sigma_{\hat{y}_i}}_{\!i} & 1-\braket{\sigma_{\hat{z}_i}}_{\!i}
\end{array}\right]\label{rhospinamat}
\end{align}
with $\braket{\sigma_{\hat{r}_i}}_{\!i}=\text{Tr}_i(\sigma_{\hat{r}_i} \rho_i^\text{spin})/\text{Tr}_i(\rho_i^\text{spin})$ 
being the expectation value of the Pauli matrix $\sigma_{\hat{r}_i}$ with respect to $\rho_i^\text{spin}$ $(\hat{r}_i=\hat{x}_i,\hat{y}_i,\hat{z}_i)$,
which is also known as spin polarization in the literature.
Eq.~\eqref{rhospina} can be regarded as the theoretical foundation for the QST analysis of $\rho_i^\text{spin}$, 
which shows how $\rho_i^\text{spin}$ 
can be reconstructed from spin polarizations $\braket{\sigma_{\hat{x}_i}}_i$, 
$\braket{\sigma_{\hat{y}_i}}_i$, and $\braket{\sigma_{\hat{z}_i}}_i$.

\subsection{Concurrence}
\label{Concur}

Several entanglement measures have been adopted to study nuclear entanglement, including entanglement entropy \cite{Nielsen:2010,Benenti:2019}, entanglement power \cite{Zanardi:2001}, negativity \cite{Vidal:2002}, and concurrence \cite{Hill:1997}.
%Among them, concurrence is widely adopted by many authors.
For the spin wave function $\ket{\Psi^\text{spin}_{AB}}$ given by Eq.~\eqref{PsiABspin},
the (spin) concurrence is given by
\begin{align}
\mathcal{C}(\ket{\Psi_{AB}^\text{spin}})
&\equiv\Braket{{\Psi^\text{spin}_{AB}}^*|\sigma_{\hat{y}_A}\otimes\sigma_{\hat{y}_B}|\Psi^\text{spin}_{AB}}\label{concurrence_definition}\\
&=2|\alpha_{00}\alpha_{11}-\alpha_{01}\alpha_{10}|\label{concurrence_expression}\\
&=2\sqrt{\text{det}(\rho_A^\text{spin})}.\label{Cdet}
\end{align}
Eq.~\eqref{concurrence_definition} is widely adopted as the definition of concurrence for pure states
with ``$*$'' being complex conjugate.
Eqs.~\eqref{concurrence_expression} and \eqref{Cdet} can be derived directly from Eq.~\eqref{concurrence_definition}
with the help of Eq.~\eqref{PsiABspin}. They will be used in later discussions.
As $2|\alpha_{00}\alpha_{11}-\alpha_{01}\alpha_{10}|\leq2|\alpha_{00}\alpha_{11}|+2|\alpha_{01}\alpha_{10}|\leq|\alpha_{00}|^2+|\alpha_{01}|^2+|\alpha_{10}|^2+|\alpha_{11}|^2=1$,
the spin concurrence satisfies
%\begin{align}
$0\leq\mathcal{C}(\ket{\Psi^\text{spin}_{AB}})\leq1$,
with
%\end{align}
$\mathcal{C}(\ket{\Psi^\text{spin}_{AB}})=0$ corresponding to the separable $\ket{\Psi^\text{spin}_{AB}}$
and 
$\mathcal{C}(\ket{\Psi^\text{spin}_{AB}})=1$ corresponding to the maximally entangled $\ket{\Psi^\text{spin}_{AB}}$.

If $\alpha_{00}$, $\alpha_{01}$, $\alpha_{10}$, and $\alpha_{11}$ are real numbers, 
the complex conjugate operator ``$*$'' makes no transformation on $\ket{\Psi^\text{spin}_{AB}}$.
In this case,
$\mathcal{C}(\ket{\Psi^\text{spin}_{AB}})$ could be reduced to the form of $\braket{\Psi^\text{spin}_{AB}|\sigma_{\hat{y}_A}\otimes\sigma_{\hat{y}_B}|\Psi^\text{spin}_{AB}}$, which is nothing but a spin correlation function with respect to $\ket{\Psi^\text{spin}_{AB}}$.
Therefore, $\mathcal{C}(\ket{\Psi^\text{spin}_{AB}})$ can be determined by measuring the expectation value of a single operator $\sigma_{\hat{y}_A}\otimes\sigma_{\hat{y}_B}$.
In the general case where $\alpha_{00}$, $\alpha_{01}$, $\alpha_{10}$, and $\alpha_{11}$ are complex numbers, 
it is shown by Refs.~\cite{Walborn:2006,Walborn:2007} that the concurrence can be determined by measuring the expectation value of a single projective operator
with respect to two copies of $\ket{\Psi^\text{spin}_{AB}}$, i.e., $\ket{\Psi^\text{spin}_{AB}}\otimes\ket{\Psi^\text{spin}_{AB}}$. This approach will be named as the Walborn protocol after the first author of Refs.~\cite{Walborn:2006,Walborn:2007}.
Explicitly, the master formula for the Walborn protocol is given by 
\begin{align}
\!\!\!\!\mathcal{C}(\ket{\Psi^\text{spin}_{AB}})
=\sqrt{\bra{\Psi^\text{spin}_{AB}}\otimes\bra{\Psi^\text{spin}_{AB}}\mathcal{W}\ket{\Psi^\text{spin}_{AB}}\otimes\ket{\Psi^\text{spin}_{AB}}},\label{concurrence_single}
\end{align}
with $\mathcal{W}\equiv4\ket{\chi}\bra{\chi}$ being a projective operator made of $\ket{\chi}=(\ket{0_A0_B1_A1_B}-\ket{0_A1_B1_A0_B}-\ket{1_A0_B0_A1_B}+\ket{1_A1_B0_A0_B})/2$ \cite{Walborn:2006,Walborn:2007}.
Although Eq.~\eqref{concurrence_single} always holds mathematically,
implementing it in different physical systems can be highly nontrivial. 
In Refs.~\cite{Walborn:2006,Walborn:2007}, an experimental setup has been designed to implement Eq.~\eqref{concurrence_single} in quantum optics,
which provides an elegant experimental probe for the amount of spin entanglement in photon pairs.
On the other hand, 
it is very difficult to implement the same equation in nuclear experiments for the following reasons.
First, given $\ket{\Psi^\text{spin}_{AB}}$ as a two-nucleon state,
 $\ket{\Psi^\text{spin}_{AB}}\otimes\ket{\Psi^\text{spin}_{AB}}$ corresponds to a four-nucleon state.
 As a result,
 the experimental determination of the expectation value $\bra{\Psi^\text{spin}_{AB}}\otimes\bra{\Psi^\text{spin}_{AB}}\mathcal{W}\ket{\Psi^\text{spin}_{AB}}\otimes\ket{\Psi^\text{spin}_{AB}}$ needs coincident measurements of spin degrees of freedom of four nucleons in the final state,
 which is a nontrivial task in nuclear physics.
 In addition, it is not clear how to implement the projection operator $\mathcal{W}=4\ket{\chi}\bra{\chi}$ in nuclear systems.

Given the difficulties in realizing the Walborn protocol in nuclear physics,
it is reasonable to consider the more conservative QST protocol.
Substituting Eq.~\eqref{rhospinamat} into Eq.~\eqref{Cdet}, the concurrence can be rewritten in terms of spin polarizations as follows:
\begin{align}
\mathcal{C}(\ket{\Psi^\text{spin}_{AB}})=\sqrt{1-\braket{\sigma_{\hat{x}_A}}^{\!2}_{\!A}-\braket{\sigma_{\hat{y}_A}}^{\!2}_{\!A}-\braket{\sigma_{\hat{z}_A}}^{\!2}_{\!A}},
\label{C_spin_pol}
\end{align}
which means that $\mathcal{C}(\ket{\Psi^\text{spin}_{AB}})$
can be determined by measuring the spin polarizations $\braket{\sigma_{\hat{x}_A}}_{\!A}$, $\braket{\sigma_{\hat{y}_A}}_{\!A}$, and $\braket{\sigma_{\hat{z}_A}}_{\!A}$ for the nucleon $A$ from the entangled nucleon pair.
The above discussion can be applied equally to the nucleon B.
Compared to the Walborn protocol, the QST protocol has the disadvantage that the expectation values of three different quantum operators have to be measured instead of one in the Walborn protocol, but has the key advantage that it can be realized easily in nuclear experiments.

\section{Experimental Implementations}
\label{SCM}

 \begin{figure}

\centering
  \includegraphics[width=0.75\linewidth]{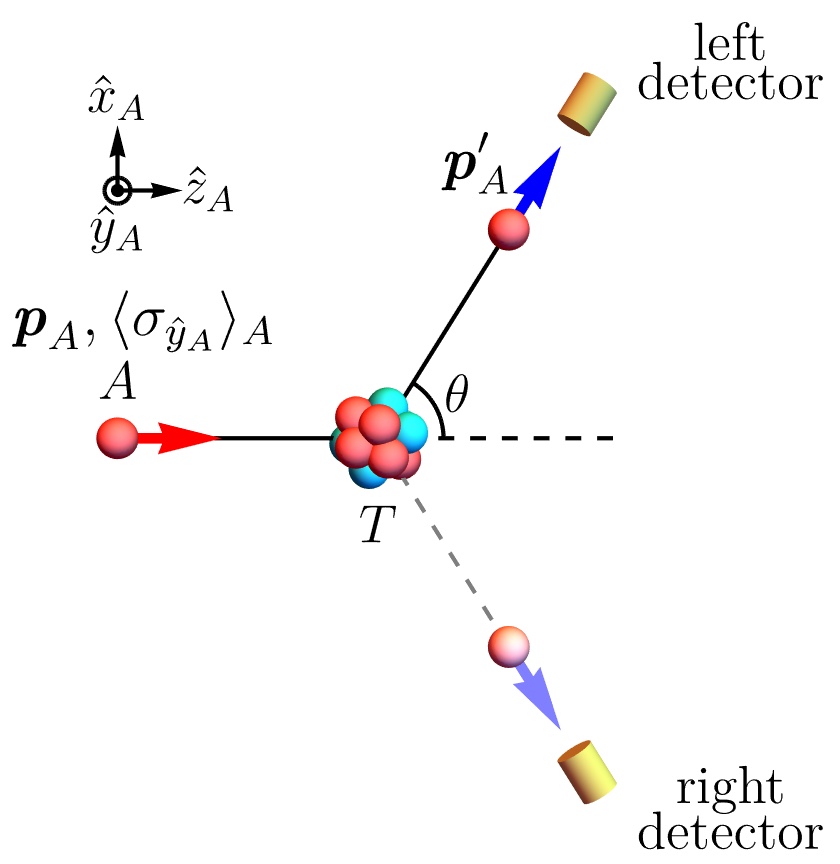}
  \caption{
The experimental setup for determining the transverse spin polarization $\braket{\sigma_{\hat{y}_A}}_A$
by the scattering between the nucleon $A$ from the entangled nucleon pair and the spin-0 nucleus $T$.
The momenta $\bm{p}_A$ and $\bm{p}_A'$ lying in the $\hat{x}_A$-$\hat{z}_A$ plane are the initial and final momenta before and after the scattering.
$\braket{\sigma_{\hat{y}_A}}_A$ could be determined by the left-right asymmetry measurement. 
The other transverse spin polarization $\braket{\sigma_{\hat{x}_A}}_A$
can be determined by using a similar experimental setup, with the final momentum $\bm{p}_A'$ selected instead in the $\hat{y}_A$-$\hat{z}_A$ plane.
}

  \label{single_scattering}
       
\end{figure}

 \begin{figure}

\centering
  \includegraphics[width=0.75\linewidth]{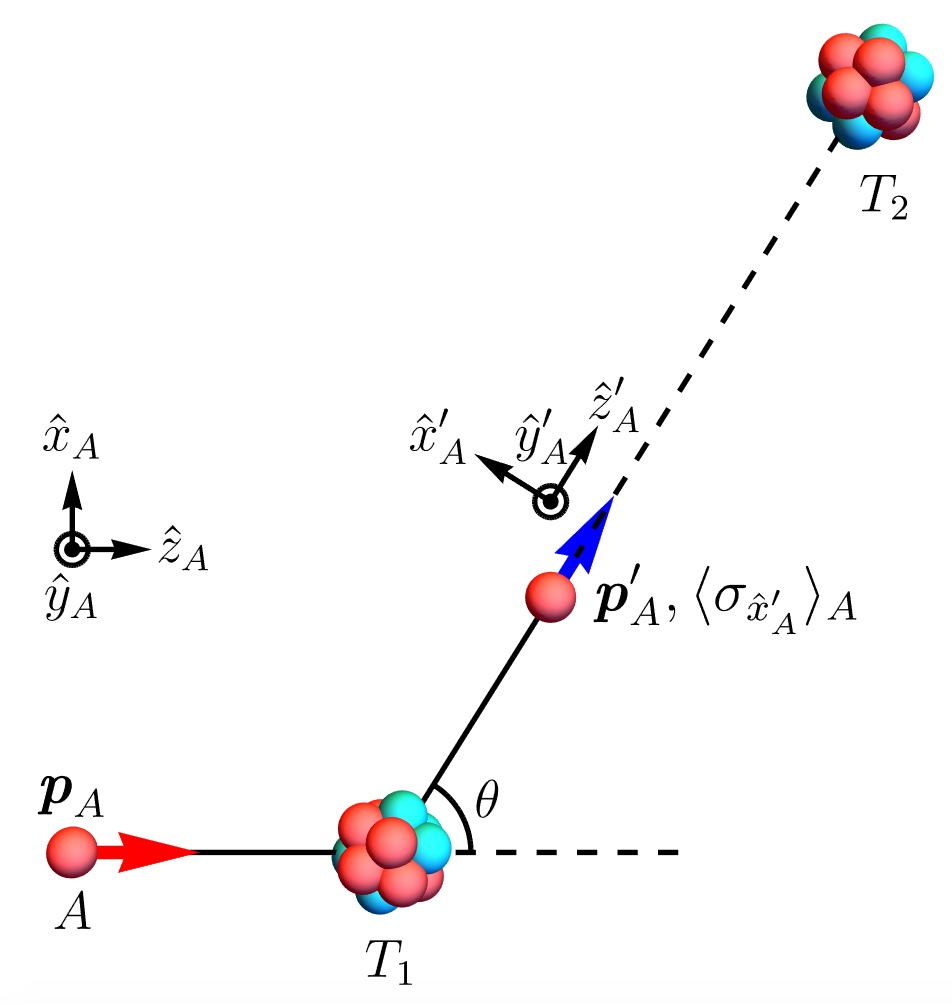}
    \includegraphics[width=0.75\linewidth]{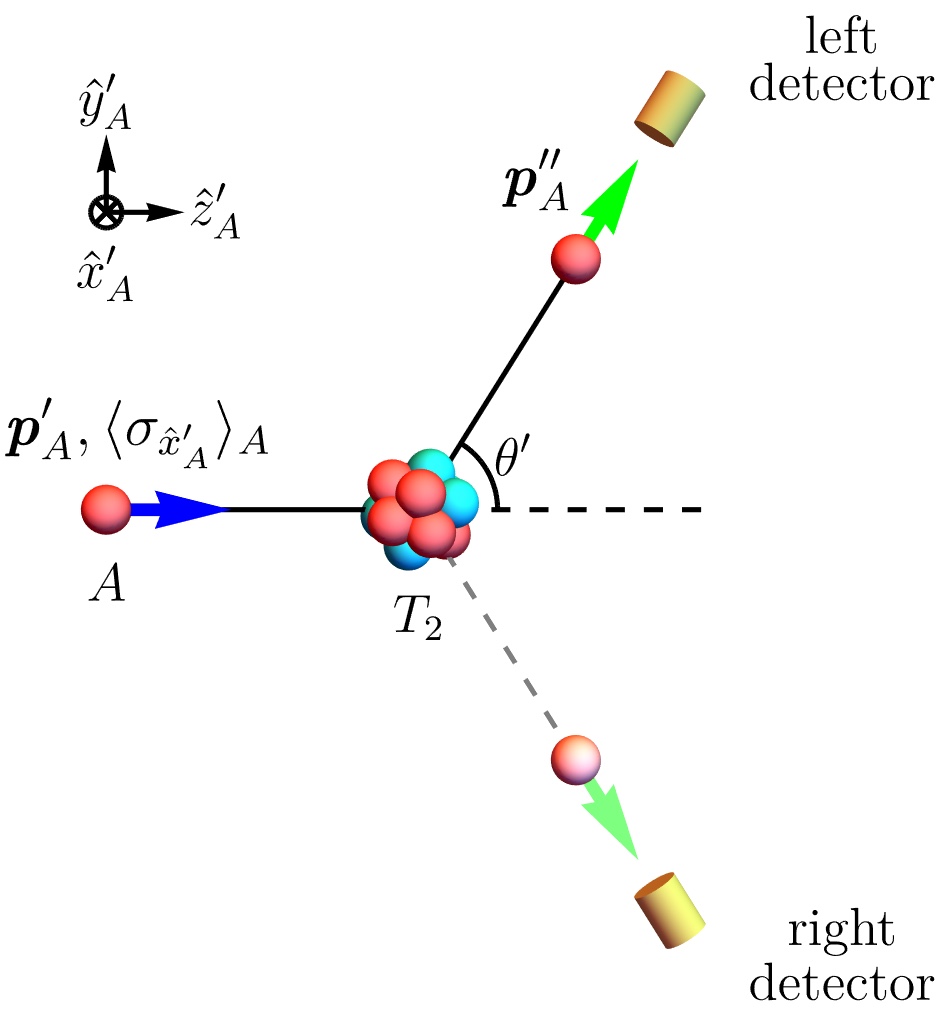}
  \caption{
The experimental setup for determining the longitudinal spin polarization $\braket{\sigma_{\hat{z}_A}}_A$
by a double scattering of the nucleon $A$ against two spin-0 nuclei $T_1$ and $T_2$.
In the first scattering given by the upper part of this figure, the initial and final momenta $\bm{p}_A$ and $\bm{p}_A'$ lie in the $\hat{x}_A$-$\hat{z}_A$ plane.
In the second scattering given by the lower part of this figure, the initial and final momenta $\bm{p}_A'$ and $\bm{p}_A''$ lie in the $\hat{y}_A'$-$\hat{z}_A'$ plane.
%The double scattering process is drawn in two parts for the convenience of illustration.
}

  \label{double_scattering}
       
\end{figure}

In this section, how to measure spin polarizations $\braket{\sigma_{\hat{x}_A}}_A$,
$\braket{\sigma_{\hat{y}_A}}_A$, and
$\braket{\sigma_{\hat{z}_A}}_A$
experimentally is shown.
As mentioned before, once their values are determined,
the concurrence of $\ket{\Psi^\text{spin}_{AB}}$ can be obtained according to Eq.~\eqref{C_spin_pol}.
Nuclear scattering with polarized nucleons will be used heavily in the following discussions,
which is the standard method to measure spin polarizations.
Theoretical and experimental studies of polarized nucleons rose in the 1950s, got boosted in the 1960s,
and became relatively mature in the 1970s.
See Refs.~\cite{Wolfenstein:1956,MacGregor:1960,Hoshizaki:1968,Ohlsen:1972,Schieck:2012,Shen:2023,Nurushev:2012} for review articles and textbooks.
Although this direction remains to be important in, e.g., understanding the nature of nucleon forces \cite{Watanabe:2022,Watanabe:2021}, 
nowadays, it is fair to say that it is less popular than before.  
Before going to details, it is crucial to stress that
the following discussions should be understood as a proof of concept.
The primary goal is to convince the community that the amount of spin entanglement of an entangled nucleon pair is measurable in nuclear experiments at least in principle
and stimulate further discussions for even better experimental protocols.
Given the high complexity of running realistic nuclear experiments,
it is natural to expect that nontrivial efforts have to be made before turning the present protocol or the similar ones into reality. 

%See Refs.~\cite{} for recent studies on nuclear physics with polarized nucleons.
%Our study relates nuclear polarized scattering to the experimental determination of spin entanglement and may provide new opportunities to this direction.
%Neverthless, as far as we know, they have never been used for the purpose of experimental determination of spin entanglement. 

First, I would like to discuss how to measure the transverse spin polarization $\braket{\sigma_{\hat{y}_A}}_A$ of the nucleon $A$,
which is a classical problem in nuclear physics with polarized nucleons.
%Here, we just follow the standard answer
The standard method is scattering the nucleon $A$ against a spin-0 target $T$ (e.g., the $^{12}$C nucleus).
It is assumed that the nucleus $T$ is much heavier than a nucleon such that the recoil effect of $T$ could be neglected for simplicity.
The nucleon $B$ is left undetected.
See Fig.~\ref{single_scattering} for the experimental setup.
The momenta $\bm{p}_A$ and $\bm{p}_A'$ are the initial and final momenta before and after the scattering.
Noticeably, they are chosen to lie in the $\hat{x}_A$-$\hat{z}_A$ plane.
The differential cross section for the parity-conserving $A$-$T$ scattering is given by \cite{Ohlsen:1972}
\begin{align}
%\!\!
&I(\theta,\phi)\nonumber\\
=&I_0(\theta)\left[1+\mathcal{A}(\theta)\braket{\bm\sigma_A}_A\cdot\bm{n}\right]\label{DiffCS}\\
=&\left\{
\begin{array}{rl}
\!\!I_0(\theta)\left[1+\mathcal{A}(\theta)\braket{\sigma_{\hat{y}_A}}_A\right]\equiv I_L(\theta) &\ \ \text{if } \phi=0,\\
\\[-1.5ex]
\!\!I_0(\theta)\left[1-\mathcal{A}(\theta)\braket{\sigma_{\hat{y}_A}}_A\right]\equiv I_R(\theta) &\ \ \text{if } \phi=\pi,
\end{array}
\right.
\end{align}
where 
$\theta$ is the relative angle between $\bm{p}_A$ and $\bm{p}_A'$,
$\bm{n}$ is the unit vector along the direction $\bm{p}_A\times\bm{p}_A'$,
$\phi$ is the relative angle between the polarization vector $\braket{\bm\sigma_A}_A\equiv(\braket{\sigma_{\hat{x}_A}}_A,\braket{\sigma_{\hat{y}_A}}_A,\braket{\sigma_{\hat{z}_A}}_A)$ 
and the unit vector $\bm{n}$,
$I(\theta,\phi)$ is the differential cross section for the $A$-$T$ scattering,
$I_0(\theta)$ is the differential cross section obtained as if the nucleon $A$ was completely unpolarized, 
with the corresponding reduced density matrix given by $\rho^\text{spin}_A=\bm{1}_2/2$ ($\bm{1}_2$ as the $2\times2$ identity matrix),
and $\mathcal{A}(\theta)$ is the so-called analyzing power,
which could be measured priorly via the polarized nucleon-$T$ scattering.
%In Eq.~\eqref{DiffCS}, it is assumed that the nucleon-nucleus interaction respects parity invariance. 
As $\bm{p}_A$ and $\bm{p}_A'$ lie in the $\hat{x}_A$-$\hat{z}_A$ plane,
the unit vector $\bm{n}$ points to the direction of $\hat{y}_A$ if $\bm{p}_A'$ points to the left-handed side of the beam direction ($\phi=0$)
and to the opposite direction of $\hat{y}_A$ if $\bm{p}_A'$ points to the right-handed side of the beam direction $(\phi=\pi)$. 
If the analyzing power $\mathcal{A}(\theta)$ is known, 
the spin polarization $\braket{\sigma_{\hat{y}_A}}_A$ could be determined by the left-right asymmetry measurement,
\begin{align}
\braket{\sigma_{\hat{y}_A}}_A=\frac{1}{\mathcal{A}(\theta)}\frac{N_L(\theta)-N_R(\theta)}{N_L(\theta)+N_R(\theta)}.\label{LSA}
\end{align}
Here, $N_L(\theta)$ and $N_R(\theta)$ are the numbers of events counted by left and right detectors,
which are proportional to the differential cross sections $I_L(\theta)$ and $I_R(\theta)$.
The other transverse spin polarization $\braket{\sigma_{\hat{x}_A}}_A$ can be measured in a similar way by
selecting the final momentum $\bm{p}_A'$ in the $\hat{y}_A$-$\hat{z}_A$ plane.
In this case, the vector $\bm{n}$ points to the direction of $\hat{x}_A$ or the opposite,
and the scalar product $\braket{\bm{\sigma}_A}_A\cdot\bm{n}$ in Eq.~\eqref{DiffCS} is reduced to 
$\pm\braket{\sigma_{\hat{x}_A}}_A$,
whose value could be determined by the left-right asymmetry measurement in the  $\hat{y}_A$-$\hat{z}_A$ plane.
 
Then, I would like to discuss how to measure the longitudinal spin polarization $\braket{\sigma_{\hat{z}_A}}_A$ of the nucleon $A$.
This can be done by a double scattering experiment, 
combined with transverse spin polarizations $\braket{\sigma_{\hat{x}_A}}_A$ and $\braket{\sigma_{\hat{y}_A}}_A$ obtained in the previous paragraph.
The corresponding experimental setup is given by Fig.~\ref{double_scattering}
with the upper part showing the first scattering between the nucleon $A$ and the spin-0 nucleus $T_1$
and the lower part showing the second scattering between the nucleon $A$ and another spin-0 nucleus $T_2$.
The double scattering process is drawn in two parts for the convenience of illustration.
%They should be understood as taking place sequentially.
In the first scattering,
the initial and final momenta are $\bm{p}_A$ and $\bm{p}_A'$
lying in the $\hat{x}_A$-$\hat{z}_A$ plane.
For the parity-conserving scattering, one has \cite{Ohlsen:1972}
\begin{align}
&\!\!\!\!I(\theta)=I_0(\theta)\left[1+\braket{\sigma_{\hat{y}_A}}_A\mathcal{A}(\theta)\right],\label{DiffCSDS1}\\
&\!\!\!\!I(\theta)=\frac{I_0(\theta)}{\braket{\sigma_{\hat{x}_A'}}_A}\left[\braket{\sigma_{\hat{x}_A}}_A\mathcal{K}_{\hat{x}_A}^{\hat{x}_A'}(\theta)+\braket{\sigma_{\hat{z}_A}}_A\mathcal{K}_{\hat{z}_A}^{\hat{x}_A'}(\theta)\right],\label{DiffCSDS2}
\end{align}
where $I(\theta)$ is the differential cross section of the $A$-$T_1$ scattering,
$I_0(\theta)$ is the unpolarized differential cross section obtained as if the nucleon $A$
is completely unpolarized,
$\braket{\sigma_{\hat{x}_A}}_A$, $\braket{\sigma_{\hat{y}_A}}_A$, and $\braket{\sigma_{\hat{z}_A}}_A$
are the spin polarizations in the initial state along the directions of $\hat{x}_A$, $\hat{y}_A$, and $\hat{z}_A$,
$\braket{\sigma_{\hat{x}_A}}_A$ is the spin polarization in the final state along the direction of $\hat{x}_A'=\bm{p}_A'/|\bm{p}_A'|$,
$\mathcal{A}(\theta)$ is the analyzing power, 
$\mathcal{K}_{\hat{x}_A}^{\hat{x}_A'}(\theta)$ and $\mathcal{K}_{\hat{z}_A}^{\hat{x}_A'}(\theta)$ are two polarization transfer coefficients. 
Substituting Eq.~\eqref{DiffCSDS1} into Eq.~\eqref{DiffCSDS2} to eliminate $I(\theta)$ and $I_0(\theta)$, one has
\begin{align}
&\braket{\sigma_{\hat{z}_A}}_A\nonumber\\
=&\frac{\braket{\sigma_{\hat{x}_A'}}_A\left[1+\braket{\sigma_{\hat{y}_A}}_A\mathcal{A}(\theta)\right]
-\braket{\sigma_{\hat{x}_A}}_A\mathcal{K}_{\hat{x}_A}^{\hat{x}_A'}(\theta)}{\mathcal{K}_{\hat{z}_A}^{\hat{x}_A'}(\theta)}.\label{LongSpinPol}
\end{align}
On the right-hand side of Eq.~\eqref{LongSpinPol}, the transverse spin polarizations $\braket{\sigma_{\hat{x}_A}}_A$ and $\braket{\sigma_{\hat{y}_A}}_A$ can be measured by following the experimental procedure presented in the previous paragraph,
while the values of $\mathcal{A}(\theta)$, $\mathcal{K}_{\hat{x}_A}^{\hat{x}_A'}(\theta)$ and $\mathcal{K}_{\hat{z}_A}^{\hat{x}_A'}(\theta)$ can be taken from other experimental studies.
Therefore, the longitudinal spin polarization $\braket{\sigma_{\hat{z}_A}}_A$ can be worked out once $\braket{\sigma_{\hat{x}_A'}}_A$ is known.
This is done by scattering the outgoing nucleon in the first step to the second spin-0 nucleus $T_2$ as shown by the lower part of Fig.~\ref{double_scattering},
where the initial momentum becomes $\bm{p}_A'$ and the final one is denoted by $\bm{p}_A''$.
It is crucial to stress that $\bm{p}_A'$ and $\bm{p}_A''$ are chosen to be in the $\hat{y}_A'$-$\hat{z}_A'$ plane
such as $\braket{\sigma_{\hat{x}_A'}}_A$ can be determined by the left-right asymmetry measurement similar to that given by Eq.~\eqref{LSA},
\begin{align}
\braket{\sigma_{\hat{x}_A'}}_A=\frac{1}{\mathcal{A}'(\theta')}\frac{N_L'(\theta')-N_R'(\theta')}{N_L'(\theta')+N_R'(\theta')},\label{LSA}
\end{align}
where $\theta'$, $\mathcal{A}'$, $N_L'$, $N_R'$ are the scattering angle, the analyzing power, and the numbers of events counted by left and right detectors in the second scattering.
This completes the experimental measurement of $\braket{\sigma_{\hat{z}_A}}_A$,
and the concurrence $\mathcal{C}(\ket{\Psi^\text{spin}_{AB}})$ can be determined correspondingly according to Eq.~\eqref{C_spin_pol}.

I would also like to discuss the connections and differences between 
the present work and Ref.\ \cite{Lamehi-Rachti:1976} (also Refs.\ \cite{Sakai:2006,Polachic:2004}).
As mentioned before,
spin correlations are measured for the two-proton system in Ref.\ \cite{Lamehi-Rachti:1976} to test Bell's inequalities.
As all entangled pure states violate a Bell's inequality \cite{Gisin:1991},
 the present work is related to Ref.\ \cite{Lamehi-Rachti:1976} in the sense that both works 
lie on entangled spin wave functions.
However, these works have different goals.
While Ref.\ \cite{Lamehi-Rachti:1976} is aimed at testing Bell's inequalities by using entangled spin wave functions produced by the proton-proton scattering,
the present work is aimed towards an experimental protocol to determine the concurrence of general spin wave functions of nucleon pairs.
Moreover, it is the spin correlations $\braket{{\bm\sigma}_A\!\cdot\!{\bm n_A}\,{\bm\sigma}_B\!\cdot\!{\bm n_B}}$ that are measured experimentally in Ref.\ \cite{Lamehi-Rachti:1976},
with $\bm{n}_A$ and $\bm{n}_B$ being the spin orientations of the two nucleons $A$ and $B$.
In contrast, it is the three spin polarizations $\{\braket{\sigma_{\hat{x}_A}}_A,\braket{\sigma_{\hat{y}_A}}_A,\braket{\sigma_{\hat{z}_A}}_A\}$ of one nucleon (say, the nucleon $A$) that are proposed to be measured in the present work.
Therefore, different spin observables are dealt with in these two works.

%The present work is different from Ref.\ \cite{Lamehi-Rachti:1976} in the following ways.
%While Ref.\ \cite{Lamehi-Rachti:1976} is aimed at doing a nuclear Bell test, 
%the present work is aimed at proposing an experimental protocol to determine the amount of spin entanglement of nucleon pairs. 
%Bell's inequality violation and quantum entanglement are closely related to each other. 

\section{Errors From Imperfect Preparation}

In previous sections, it is assumed that the spin component of the entangled nucleon pair is prepared in a pure state $\ket{\Psi^\text{spin}_{AB}}$.
However, in realistic nuclear experiments, it is difficult to fulfill this condition perfectly.
In general, the entangled nucleon pair is often prepared in a mixed state,
which, without loss of generality,
is chosen to be
\begin{align}
\bar{\rho}_{AB}^\text{spin}=(1-\varepsilon)\ket{\Psi^\text{spin}_{AB}}\bra{\Psi^\text{spin}_{AB}}+\varepsilon \ket{\tilde{\Psi}^\text{spin}_{AB}}\bra{\tilde{\Psi}^\text{spin}_{AB}},
\end{align}
where $\ket{\tilde{\Psi}_{AB}^\text{spin}}$ is the impurity spin wave function mixed with $\ket{\Psi^\text{spin}_{AB}}$ and  $\varepsilon\ll 1$ is the probability of $\ket{\tilde{\Psi}_{AB}^\text{spin}}$ in $\bar{\rho}_{AB}^\text{spin}$.
It is important to estimate the error caused by mistaking $\bar{\rho}_{AB}^\text{spin}$ for the idealized pure state $\ket{\Psi^\text{spin}_{AB}}\bra{\Psi^\text{spin}_{AB}}$.
In this case, the realistic spin polarization of the nucleon $A$ is found to be
\begin{align}
\braket{\bar{\sigma}_{\hat{r}_A}}_A
=(1-\varepsilon)\braket{\sigma_{\hat{r}_A}}_A+\varepsilon\braket{\tilde{\sigma}_{\hat{r}_A}}_A
\end{align}
with $\hat{r}_A\equiv\hat{x}_A,\hat{y}_A,\hat{z}_A$,  $\braket{\bar{\sigma}_{\hat{r}_A}}_A\equiv{\text{Tr}(\sigma_{\hat{r}_A}\bar{\rho}_{AB}^\text{spin})}/{\text{Tr}(\bar{\rho}_{AB}^\text{spin})}$,
and $\braket{\tilde{\sigma}_{\hat{r}_A}}_A\equiv{\text{Tr}(\sigma_{\hat{r}_A}\tilde{\rho}_{AB}^\text{spin})}/{\text{Tr}(\tilde{\rho}_{AB}^\text{spin})}$.
It is then straightforward to see that the concurrence ${\mathcal{C}}(\bar{\rho}^\text{spin}_{AB})$ derived from $\braket{\bar{\sigma}_{\hat{r}_A}}_A$ by Eq.~\eqref{C_spin_pol} deviates from the target ${\mathcal{C}}(\ket{{\Psi}^\text{spin}_{AB}})$ by a small error proportional to $\varepsilon$,
\begin{align}
\mathcal{C}(\bar{\rho}^\text{spin}_{AB})=\mathcal{C}(\ket{{\Psi}^\text{spin}_{AB}})+\mathcal{O}(\varepsilon).
\end{align}
In other words, the error caused by imperfect preparation of nucleon pairs is roughly proportional to the probability of the impurity state.

\section{Conclusions}
\label{Concl}

Nuclear entanglement has attracted much attention from the theoretical side.
In this work, the problem of how to measure it in nuclear experiments is explored.
Explicitly, the experimental determination of the amount of spin entanglement of nucleon pairs
is studied under the inspiration of QST.
Although not the unique approach,
it is argued that QST is more suitable for nuclear physics than, e.g., the Walborn protocol.
It is shown that the concurrence of the spin wave function can be determined by measuring three spin polarizations of one nucleon from the entangled pair
via several polarization experiments.
The errors caused by imperfect preparation of the entangled nucleon pair are also discussed.
%This work makes a preliminary attempt to relate theoretical studies of nuclear entanglement to experimental 

This work could be improved and generalized in several directions.
First,
the present study should be regarded as a proof of concept rather than a detailed experimental plan,
and it is important to go deeper into the experimental details of the measurement protocol described in Sec.~\ref{SCM}.
This calls for further cooperation between theoretical and experimental sides of nuclear physics.
Second, it seems feasible to generalize the discussions to entangled states made of nuclei with spin-$1/2$ and higher spins. 
These states could be produced naturally in nucleus-nucleus scattering and reaction processes.
Third, it is interesting to explore how to measure the amount of entanglement in forms other than spin entanglement.
In several nuclear structure studies, orbital entanglement is calculated for a couple of nuclei.
It is interesting to explore the experimental determination of orbital entanglement as well.
If worked out, it will promote the research of nuclear entanglement significantly. 
Last but not least, 
it would also be interesting to generalize the present work to more than two nucleons.
Several multipartite entanglement measures have been proposed in quantum information science, such as
$n$-tangle \cite{Coffman:2000,Wong:2001}, multipartite concurrence \cite{Carvalho:2004,Aolita:2006}, and concurrence fill \cite{Xie:2021,Xie:2023dzw}.
Noticeably, $n$-tangle has been adopted in Ref.~\cite{Hengstenberg:2023ryt} to study multipartite entanglement and information rearrangement in nuclear many-body systems.
Nevertheless, these problems are beyond my current scope
and are left for future studies.

\begin{acknowledgments}
This work is supported by the National Natural Science Foundation of China (Grant No.\ 12375122)
and the Fundamental Research Funds for the Central Universities (Grant No.\ B230201022).
\end{acknowledgments}


\begin{thebibliography}{999}

%\bibitem{Fei:2009}
%S.-M.\ Fei, M.-J.\ Zhao, K.\ Chen, and Z.-X.\ Wang, 
%Phys.\ Rev.\ A {\bf 80}, 032320 (2009).

%\bibitem{Verstraete:2002}
%F.\ Verstraete and M. M. Wolf, 
%Phys.\ Rev.\ Lett.\ {\bf 89}, 170401 (2002).

%%\cite{Bai:2022hfv}
%\bibitem{Bai:2022hfv}
%D.~Bai and Z.~Ren,
%%``Entanglement generation in few-nucleon scattering,''
%Phys.\ Rev.\ C \textbf{106}, 064005 (2022).
%%doi:10.1103/PhysRevC.106.064005
%%[arXiv:2212.11092 [nucl-th]].
%%5 citations counted in INSPIRE as of 16 Jun 2023
%
%%\cite{Bai:2023rkc}
%\bibitem{Bai:2023rkc}
%D.~Bai,
%%``Quantum information in nucleon-nucleon scattering,''
%Phys.\ Rev.\ C \textbf{107}, 044005 (2023).
%%doi:10.1103/PhysRevC.107.044005
%%1 citations counted in INSPIRE as of 16 Jun 2023
%
%%\cite{Bai:2023tey}
%\bibitem{Bai:2023tey}
%D.~Bai,
%%``Spin entanglement in neutron-proton scattering,''
%arXiv:2306.04918 [nucl-th].
%%1 citations counted in INSPIRE as of 16 Jun 2023

%\cite{Kwasniewicz:2013cqa,Kanada-Enyo:2015ncq,Kanada-Enyo:2015kyo,Legeza:2015fja,Gorton:2018,Johnson:2019,Kruppa:2020rfa,Kovacs:2021yme,Kwasniewicz:2016,Kwasniewicz:2017dbc,Gorton:2018,Johnson:2019,Beane:2018oxh,Robin:2020aeh,Low:2021ufv,Beane:2021zvo,Jafarizadeh:2022kcq,Kovacs:2021yme,Pazy:2022mmg,Tichai:2022bxr,Bai:2022hfv,Liu:2022grf,Johnson:2022mzk,Bai:2023rkc,Gu:2023aoc}
\bibitem{Kwasniewicz:2013cqa}
E.~Kwa\'sniewicz,
%``Entanglement of 1s0d-shell nucleon pairs,''
J.\ Phys.\ G \textbf{41}, 015107 (2014).
%doi:10.1088/0954-3899/41/1/015107
%3 citations counted in INSPIRE as of 20 Apr 2022

%\cite{Kanada-Enyo:2015ncq,Kanada-Enyo:2015kyo}
\bibitem{Kanada-Enyo:2015ncq}
Y.~Kanada-En'yo,
%``Analysis of delocalization of clusters in linear-chain $\alpha$-cluster states with entanglement entropy,''
Phys.\ Rev.\ C \textbf{91}, 034303 (2015).
%doi:10.1103/PhysRevC.91.034303
%[arXiv:1501.06231 [nucl-th]].
%5 citations counted in INSPIRE as of 20 Apr 2022

%\cite{Kanada-Enyo:2015kyo}
\bibitem{Kanada-Enyo:2015kyo}
Y.~Kanada-En'yo,
%``Entanglement entropy and Schmidt number as measures of delocalization of $\alpha$ clusters in one-dimensional nuclear systems,''
Prog.\ Theor.\ Exp.\ Phys.\ \textbf{2015}, 043D04 (2015).
%doi:10.1093/ptep/ptv050
%[arXiv:1501.06639 [nucl-th]].
%2 citations counted in INSPIRE as of 20 Apr 2022

%\cite{Legeza:2015fja,Gorton:2018,Johnson:2019,Kruppa:2020rfa,Kovacs:2021yme}
\bibitem{Legeza:2015fja}
\"O.~Legeza, L.~Veis, A.~Poves, and J.~Dukelsky,
%``Advanced density matrix renormalization group method for nuclear structure calculations,''
Phys.\ Rev.\ C \textbf{92}, 051303(R) (2015).
%doi:10.1103/PhysRevC.92.051303
%[arXiv:1507.00161 [nucl-th]].
%16 citations counted in INSPIRE as of 20 Apr 2022

\bibitem{Kwasniewicz:2016}
E.~Kwa\'sniewicz,
Acta Phys.\ Pol.\ B {\bf 47}, 2297 (2016).

%\cite{Kwasniewicz:2017dbc}
\bibitem{Kwasniewicz:2017dbc}
E.~Kwa\'sniewicz and D.~Kurzyk,
%``Entanglement of 0p-, 1s0d- and 1p0f-shell nucleon pairs,''
Int.\ J.\ Mod.\ Phys.\ E \textbf{26}, 1750023 (2017).
%doi:10.1142/S0218301317500239
%1 citations counted in INSPIRE as of 20 Apr 2022

\bibitem{Gorton:2018}
O.\ C.\ Gorton, \emph{Efficient modeling of nuclei through coupling of proton and neutron wavefunctions},
Master’s thesis, San Diego State University, 2018 (unpublished).

\bibitem{Johnson:2019}
C.\ W.\ Johnson, \emph{Entanglement entropy and proton-neutron
 interactions}, presented at the ESNT
Workshop on Proton-Neutron Pairing, 2019 (unpublished).
%http://esnt.cea.fr/Phocea/Page/index.php?id=84

%\cite{Beane:2018oxh}
\bibitem{Beane:2018oxh}
S.~R.~Beane, D.~B.~Kaplan, N.~Klco, and M.~J.~Savage,
%``Entanglement Suppression and Emergent Symmetries of Strong Interactions,''
Phys.\ Rev.\ Lett.\ \textbf{122}, 102001 (2019).
%doi:10.1103/PhysRevLett.122.102001
%[arXiv:1812.03138 [nucl-th]].
%27 citations counted in INSPIRE as of 20 Apr 2022

%\cite{Kruppa:2020rfa}
\bibitem{Kruppa:2020rfa}
A.~T.~Kruppa, J.~Kov\'acs, P.~Salamon, and \"O.~Legeza,
%``Entanglement and correlation in two-nucleon systems,''
J.\ Phys.\ G \textbf{48}, 025107 (2021).
%doi:10.1088/1361-6471/abc2dd
%[arXiv:2006.07448 [nucl-th]].
%5 citations counted in INSPIRE as of 20 Apr 2022

%\cite{Robin:2020aeh}
\bibitem{Robin:2020aeh}
C.~Robin, M.~J.~Savage, and N.~Pillet,
%``Entanglement Rearrangement in Self-Consistent Nuclear Structure Calculations,''
Phys.\ Rev.\ C \textbf{103}, 034325 (2021).
%doi:10.1103/PhysRevC.103.034325
%[arXiv:2007.09157 [nucl-th]].
%17 citations counted in INSPIRE as of 20 Apr 2022

%\cite{Low:2021ufv,Beane:2021zvo,Jafarizadeh:2022kcq,Kovacs:2021yme,Pazy:2022mmg,Tichai:2022bxr,Bai:2022hfv,Liu:2022grf,Johnson:2022mzk,Bai:2023rkc,Gu:2023aoc}
\bibitem{Low:2021ufv}
I.~Low and T.~Mehen,
%``Symmetry from entanglement suppression,''
Phys.\ Rev.\ D \textbf{104}, 074014 (2021).
%doi:10.1103/PhysRevD.104.074014
%[arXiv:2104.10835 [hep-th]].
%6 citations counted in INSPIRE as of 20 Apr 2022

%\cite{Beane:2021zvo}
\bibitem{Beane:2021zvo}
S.~R.~Beane, R.~C.~Farrell, and M.~Varma,
%``Entanglement minimization in hadronic scattering with pions,''
Int.\ J.\ Mod.\ Phys. A \textbf{36}, 2150205 (2021).
%doi:10.1142/S0217751X21502055
%[arXiv:2108.00646 [hep-ph]].
%3 citations counted in INSPIRE as of 20 Apr 2022

%\cite{Jafarizadeh:2022kcq}
\bibitem{Jafarizadeh:2022kcq}
M.~A.~Jafarizadeh, M.~Ghapanvari, and N.~Amiri,
%``Entanglement entropy as a signature of a quantum phase transition in nuclei in the framework of the interacting boson model and interacting boson-fermion model,''
Phys.\ Rev.\ C \textbf{105}, 014307 (2022).
%doi:10.1103/PhysRevC.105.014307
%0 citations counted in INSPIRE as of 09 Nov 2022

%\cite{Kovacs:2021yme}
\bibitem{Kovacs:2021yme}
A.~T.~Kruppa, J.~Kov\'acs, P.~Salamon, \"O.~Legeza, and G.~Zar\'and,
%``Entanglement and seniority,''
Phys.\ Rev.\ C {\bf 106}, 024303 (2022).
%arXiv:2112.15513 [nucl-th].
%0 citations counted in INSPIRE as of 20 Apr 2022

%\cite{Bulgac:2022cjg,Bulgac:2022ygo}
\bibitem{Bulgac:2022cjg}
A.~Bulgac, M.~Kafker, and I.~Abdurrahman,
%``Measures of complexity and entanglement in many-fermion systems,''
Phys.\ Rev.\ C \textbf{107}, 044318 (2023).
%doi:10.1103/PhysRevC.107.044318
%[arXiv:2203.04843 [nucl-th]].
%11 citations counted in INSPIRE as of 30 Jul 2023

%\cite{Bulgac:2022ygo}
\bibitem{Bulgac:2022ygo}
A.~Bulgac,
%``Entanglement entropy, single-particle occupation probabilities, and short-range correlations,''
Phys.\ Rev.\ C \textbf{107}, L061602 (2023).
%doi:10.1103/PhysRevC.107.L061602
%[arXiv:2203.12079 [nucl-th]].
%12 citations counted in INSPIRE as of 30 Jul 2023

%\cite{Pazy:2022mmg}
\bibitem{Pazy:2022mmg}
E.~Pazy,
%``Entanglement entropy between short range correlations and the Fermi sea in nuclear structure,''
Phys.\ Rev.\ C \textbf{107}, 054308 (2023).
%doi:10.1103/PhysRevC.107.054308
%[arXiv:2206.10702 [nucl-th]].
%9 citations counted in INSPIRE as of 30 Jul 2023

%\cite{Tichai:2022bxr}
\bibitem{Tichai:2022bxr}
A.~Tichai, S.~Knecht, A.~T.~Kruppa, \"O.~Legeza, C.~P.~Moca, A.~Schwenk, M.~A.~Werner, and G.~Zarand,
%``Combining the in-medium similarity renormalization group with the density matrix renormalization group: Shell structure and information entropy,''
Phys.\ Lett.\ B \textbf{845}, 138139 (2023).
%doi:10.1016/j.physletb.2023.138139
%[arXiv:2207.01438 [nucl-th]].
%16 citations counted in INSPIRE as of 03 Feb 2024

%\cite{Bai:2022hfv}
\bibitem{Bai:2022hfv}
D.~Bai and Z.~Ren,
%``Entanglement generation in few-nucleon scattering,''
Phys.\ Rev.\ C \textbf{106}, 064005 (2022).
%doi:10.1103/PhysRevC.106.064005
%[arXiv:2212.11092 [nucl-th]].
%3 citations counted in INSPIRE as of 03 May 2023

%\cite{Liu:2022grf}
\bibitem{Liu:2022grf}
Q.~Liu, I.~Low, and T.~Mehen,
%``Minimal entanglement and emergent symmetries in low-energy QCD,''
Phys.\ Rev.\ C \textbf{107}, 025204 (2023).
%doi:10.1103/PhysRevC.107.025204
%[arXiv:2210.12085 [quant-ph]].
%3 citations counted in INSPIRE as of 03 May 2023

%\cite{Johnson:2022mzk}
\bibitem{Johnson:2022mzk}
C.~W.~Johnson and O.~C.~Gorton,
%``Proton-neutron entanglement in the nuclear shell model,''
J.\ Phys.\ G \textbf{50}, 045110 (2023).
%doi:10.1088/1361-6471/acbece
%[arXiv:2210.14338 [nucl-th]].
%4 citations counted in INSPIRE as of 03 May 2023

%\cite{Bai:2023rkc}
\bibitem{Bai:2023rkc}
D.~Bai,
%``Quantum information in nucleon-nucleon scattering,''
Phys.\ Rev.\ C \textbf{107}, 044005 (2023).
%doi:10.1103/PhysRevC.107.044005
%0 citations counted in INSPIRE as of 03 May 2023

%\cite{Gu:2023aoc}
\bibitem{Gu:2023aoc}
C.~Gu, Z.~H.~Sun, G.~Hagen, and T.~Papenbrock,
%``Entanglement entropy of nuclear systems,''
Phys.\ Rev.\ C \textbf{108}, 054309 (2023).
%doi:10.1103/PhysRevC.108.054309
%[arXiv:2303.04799 [nucl-th]].
%12 citations counted in INSPIRE as of 30 Jan 2024

%\cite{Sun:2023nzj}
\bibitem{Sun:2023nzj}
Z.~H.~Sun, G.~Hagen, and T.~Papenbrock,
%``Coupled-cluster theory for strong entanglement in nuclei,''
Phys.\ Rev.\ C \textbf{108}, 014307 (2023).
%doi:10.1103/PhysRevC.108.014307
%[arXiv:2305.07577 [nucl-th]].
%1 citations counted in INSPIRE as of 30 Jul 2023

%\cite{Bai:2023tey}
\bibitem{Bai:2023tey}
D.~Bai,
%``Spin entanglement in neutron-proton scattering,''
Phys.\ Lett.\ B \textbf{845}, 138162 (2023).
%doi:10.1016/j.physletb.2023.138162
%[arXiv:2306.04918 [nucl-th]].
%6 citations counted in INSPIRE as of 30 Jan 2024

%\cite{Sun:2023nzj,Bai:2023tey,Miller:2023ujx,Miller:2023snw,Hengstenberg:2023ryt,Perez-Obiol:2023wdz}
\bibitem{Miller:2023ujx}
G.~A.~Miller,
%``Entanglement Maximization in Low-Energy Neutron-Proton Scattering,''
Phys.\ Rev.\ C \textbf{108}, L031002 (2023).
%doi:10.1103/PhysRevC.108.L031002
%[arXiv:2306.03239 [nucl-th]].
%7 citations counted in INSPIRE as of 30 Jan 2024

%\cite{Miller:2023snw}
\bibitem{Miller:2023snw}
G.~A.~Miller,
%``The Entanglement of Elastic and Inelastic Scattering,''
Phys.\ Rev.\ C \textbf{108}, L041601 (2023).
%doi:10.1103/PhysRevC.108.L041601
%[arXiv:2306.14800 [nucl-th]].
%6 citations counted in INSPIRE as of 30 Jan 2024

%\cite{Hengstenberg:2023ryt}
\bibitem{Hengstenberg:2023ryt}
S.~M.~Hengstenberg, C.~E.~P.~Robin, and M.~J.~Savage,
%``Multi-Body Entanglement and Information Rearrangement in Nuclear Many-Body Systems,''
Eur.\ Phys.\ J.\ A \textbf{59}, 231 (2023).
%doi:10.1140/epja/s10050-023-01145-x
%[arXiv:2306.16535 [nucl-th]].
%6 citations counted in INSPIRE as of 30 Jan 2024

%\cite{Perez-Obiol:2023wdz}
\bibitem{Perez-Obiol:2023wdz}
A.~P\'erez-Obiol, S.~Masot-Llima, A.~M.~Romero, J.~Men\'endez, A.~Rios, A.~Garc\'\i{}a-S\'aez, and B.~Juli\'a-D\'\i{}az,
%``Quantum entanglement patterns in the structure of atomic nuclei within the nuclear shell model,''
Eur.\ Phys.\ J.\ A \textbf{59}, 240 (2023).
%doi:10.1140/epja/s10050-023-01151-z
%[arXiv:2307.05197 [nucl-th]].
%4 citations counted in INSPIRE as of 30 Jan 2024

%\cite{Kou:2023knx}
\bibitem{Kou:2023knx}
W.~Kou, J.~Chen, and X.~Chen,
%``Exploring Short-Range Correlations in symmetric nuclei: Insights into contacts and entanglement entropy,''
Phys.\ Lett.\ B \textbf{849}, 138453 (2024).
%doi:10.1016/j.physletb.2024.138453
%[arXiv:2309.05909 [nucl-th]].
%0 citations counted in INSPIRE as of 03 Feb 2024

\bibitem{Kirchner:2023}
%\cite{Kirchner:2023dvg}
%\bibitem{Kirchner:2023dvg}
T.~Kirchner, W.~Elkamhawy, and H.~W.~Hammer,
%``Entanglement in few-nucleon scattering events,''
arXiv:2312.14484 [nucl-th].
%0 citations counted in INSPIRE as of 30 Jan 2024

\bibitem{Lamehi-Rachti:1976}
M.\ Lamehi-Rachti and W.\ Mittig,
Phys.\ Rev.\ D {\bf 14}, 2543 (1976).

\bibitem{Sakai:2006}
H.\ Sakai, T.\ Saito, T.\ Ikeda, K.\ Itoh, T.\ Kawabata, H.\ Kuboki, Y.\ Maeda, N.\ Matsui, C.\ Rangacharyulu, M.\ Sasano,
Y.\ Satou, K.\ Sekiguchi, K.\ Suda, A.\ Tamii, T.\ Uesaka, and K.\ Yako,
Phys.\ Rev.\ Lett.\ {\bf 97}, 150405 (2006)
[Erratum: Phys.\ Rev.\ Lett.\ {\bf 97}, 179901(E) (2006)].

\bibitem{Polachic:2004}
C.\ Polachic, C.\ Rangacharyulu, A.\ van den Berg, M.\ Harakeh, M.\ de Huu, H.\ Wortche, J.\ Heyse, C.\ Baumer, D.\ Frekers, and J.\ Brooke,
Phys.\ Lett.\ A {\bf 323}, 176 (2004).

%\cite{Sancho:2000,Walborn:2006,Walborn:2007,Fei:2009,Li:2013}
\bibitem{Sancho:2000}
J.\ M.\ G.\ Sancho and S.\ F.\ Huelga,
Phys.\ Rev.\ A {\bf 61}, 042303 (2000).

\bibitem{Walborn:2006}
S.\ P.\ Walborn, P.\ H.\ Souto Ribeiro, L.\ Davidovich, F.\ Mintert, and A.\ Buchleitner,
Nature {\bf 440}, 1022 (2006).

\bibitem{Walborn:2007}
S.\ P.\ Walborn, P.\ H.\ Souto Ribeiro, L.\ Davidovich, F.\ Mintert, and A.\ Buchleitner,
Phys.\ Rev.\ A {\bf 75}, 032338 (2007).

\bibitem{Fei:2009}
S.-M.\ Fei, M.-J.\ Zhao, K.\ Chen, and Z.-X.\ Wang,
Phys.\ Rev.\ A {\bf 80}, 032320 (2009).

\bibitem{Li:2013}
M.\ Li, M.-J.\ Zhao, S.-M.\ Fei, and Z.-X.\ Wang,
Front.\ Phys.\ {\bf 8}, 357 (2013).

\bibitem{Paris:2004}
M.\ Paris and J.\ \v{R}eh\'{a}\v{c}ek,
\emph{Quantum State Estimation}
(Springer-Verlag, Berlin, 2004).

\bibitem{White:1999}
A.\ G.\ White, D.\ F.\ V.\ James, P.\ H.\ Eberhard, and P.\ G.\ Kwiat,
Phys.\ Rev.\ Lett.\ {\bf 83}, 3103 (1999).

\bibitem{Nielsen:2010}
M.~A.~Nielsen and I.~L.~Chuang, 
\emph{Quantum Computation and Quantum Information - 10th Anniversary Edition} 
(Cambridge University Press, Cambridge, 2010).

%%\cite{Witten:2018zva,Headrick:2019eth}
%
%%\cite{Headrick:2019eth}
%\bibitem{Headrick:2019eth}
%M.~Headrick,
%%``Lectures on entanglement entropy in field theory and holography,''
%arXiv:1907.08126 [hep-th].
%%52 citations counted in INSPIRE as of 06 Feb 2023

\bibitem{Benenti:2019}
G.~Benenti, G.~Casati, D.~Rossini, and G.~Strini,
\emph{Principles of Quantum Computation and Information: A Comprehensive Textbook}
(World Scientific Publishing, Singapore, 2019).

\bibitem{Zanardi:2001}
P.\ Zanardi, Phys.\ Rev.\ A {\bf 63}, 040304(R) (2001).

\bibitem{Vidal:2002}
G.\ Vidal and R.\ F.\ Werner, 
Phys.\ Rev.\ A {\bf 65}, 032314
(2002).

\bibitem{Hill:1997}
S.\ A.\ Hill and W.\ K.\ Wootters,
Phys.\ Rev.\ Lett.\ {\bf 78}, 5022 (1997).

%\cite{Wolfenstein:1956,MacGregor:1960,Hoshizaki:1968,Ohlsen:1972,Taylor:1972,Schieck:2012,Shen:2023}
\bibitem{Wolfenstein:1956}
L.\ Wolfenstein,
Ann.\ Rev.\ Nucl.\ Sci.\ {\bf 6}, 43 (1956).

\bibitem{MacGregor:1960}
M.\ H.\ MacGregor, M.\ J.\ Moravcsik, and H.\ P.\ Stapp,
Ann.\ Rev.\ Nucl.\ Sci.\ {\bf 10}, 291 (1960).

\bibitem{Hoshizaki:1968}
N.\ Hoshizaki, 
Suppl.\ Prog.\ Theor.\ Phys.\ {\bf 42}, 107 (1968).

\bibitem{Ohlsen:1972}
G.\ G.\ Ohlsen, Rep.\ Prog.\ Phys.\ {\bf 35}, 717 (1972).

%\bibitem{Taylor:1972}
%J.\ R.\ Taylor,
%\emph{Scattering Theory -- The Quantum Theory of Nonrelativistic Collisions}
%(John Wiley \& Sons, New York, 1972).

\bibitem{Schieck:2012}
H.\ P.\ g.\ Schieck,
\emph{Nuclear Physics with Polarized Particles}
(Springer-Verlag, Berlin, 2012).

\bibitem{Nurushev:2012}
S.\ B.\ Nurushev, M.\ F.\ Runtso, and M.\ N.\ Strikhanov,
\emph{Introduction to Polarization Physics}
(Springer-Verlag, Berlin, 2012).

\bibitem{Shen:2023}
Q.-B.\ Shen,
\emph{Polarization Theory of Nuclear Reactions}
(Springer-Verlag, Berlin, 2023).

\bibitem{Watanabe:2021}
A.\ Watanabe, S.\ Nakai, Y.\ Wada, K.\ Sekiguchi, A.\ Deltuva, T.\ Akieda, D.\ Etoh, M.\ Inoue, Y.\ Inoue, K.\ Kawahara, H.\ Kon, K.\ Miki, T.\ Mukai, D.\ Sakai, S.\ Shibuya,
 Y.\ Shiokawa, T.\ Taguchi, H.\ Umetsu, Y.\ Utsuki, M.\ Watanabe, S.\ Goto, K.\ Hatanaka, Y.\ Hirai, T.\ Ino,
 D.\ Inomoto, A.\ Inoue, S.\ Ishikawa, M.\ Itoh, H.\ Kanda, H.\ Kasahara, N.\ Kobayashi, Y.\ Maeda, S.\ Mitsumoto, S.\ Nakamura, K.\ Nonaka, H.\ J.\ Ong, 
 H.\ Oshiro, Y.\ Otake, H.\ Sakai, A.\ Taketani, A.\ Tamii, D.\ T.\ Tran, T.\ Wakasa, Y.\ Wakabayashi, and T.\ Wakui,
 Phys.\ Rev.\ C {\bf 103}, 044001 (2021).

\bibitem{Watanabe:2022}
A.\ Watanabe, S.\ Nakai, K.\ Sekiguchi, A.\ Deltuva, S.\ Goto, K.\ Hatanaka, Y.\ Hirai, T.\ Ino, D.\ Inomoto, M.\ Inoue, S.\ Ishikawa, M.\ Itoh, H.\ Kanda,
 H.\ Kasahara, Y.\ Maeda, K.\ Miki, K.\ Nonaka, H.\ J.\ Ong, H.\ Oshiro, D.\ Sakai, H.\ Sakai, S.\ Shibuya, D.\ T.\ Tran, H.\ Umetsu, Y.\ Utsuki, and T.\ Wakasa,
 Phys.\ Rev.\ C {\bf 106}, 054002 (2022).
 
 \bibitem{Gisin:1991}
 N.\ Gisin,
 %Bell’s inequality holds for all non-product states. 
 Phys.\ Lett.\ A {\bf 154}, 201 (1991).
 
 \bibitem{Coffman:2000}
 V.\ Coffman, J.\ Kundu, and W.\ K.\ Wootters,
 Phys.\ Rev.\ A {\bf 61}, 052306 (2000).
 
 \bibitem{Wong:2001}
 A.\ Wong and N.\ Christensen, 
 Phys.\ Rev.\ A {\bf 63}, 044301 (2001).
 
 %\cite{Carvalho:2004,Aolita:2006}
 \bibitem{Carvalho:2004}
 A.\ R.\ R.\ Carvalho, F.\ Mintert, and A.\ Buchleitner,
 Phys.\ Rev.\ Lett.\ {\bf 93}, 230501 (2004).
 
 \bibitem{Aolita:2006}
 L.\ Aolita and F.\ Mintert,
 Phys.\ Rev.\ Lett.\ {\bf 97}, 050501 (2006).
 
 
 \bibitem{Xie:2021}
 S.\ Xie and J.\ H.\ Eberly, Phys.\ Rev.\ Lett.\ {\bf 127}, 040403 (2021).
 
 %\cite{Xie:2023dzw}
\bibitem{Xie:2023dzw}
S.~Xie, D.~Younis, Y.~Mei, and J.~H.~Eberly,
%``Unraveling the Mysteries of Multipartite Entanglement: A Journey Through Geometry,''
arXiv:2304.03281 [quant-ph].
%4 citations counted in INSPIRE as of 03 Feb 2024
 
% \bibitem{Bai:2024}
% D.\ Bai, \emph{Spin entanglement of multineutrons}, under preparation (2024).

\end{thebibliography}
\end{document}